Harvesting the triplet excitons of quasi-two-dimensional perovskite toward highly efficient white light-emitting diodes

*Author(s), and Corresponding Author(s)**

Yue Yu,[2,#] Chenjing Zhao,[1,#] Lin Ma,[2,#] Lihe Yan,[1] Bo Jiao,[1] Jingrui Li,[3] Jun Xi,[4] Jinhai Si,[1] Yuren Li,[1] Yanmin Xu,[1] Hua Dong,[1] Jingfei Dai,[1] Fang Yuan,[1] Peichao Zhu,[1] Alex K.-Y. Jen,[5] and Zhaoxin Wu[1]*

[1]Key Laboratory for Physical Electronics and Devices of the Ministry of Education & Shaanxi Key Lab of Information Photonic Technique, School of Electronic Science and Engineering, Xi'an Jiaotong University, Xi'an, 710049, China.

[2]School of Physics and Optoelectronic Engineering, Xidian University, Xi'an, 710071, China

[3]Electronic Materials Research Laboratory, Key Laboratory of the Ministry of Education & International Center for Dielectric Research, School of Electronic Science and Engineering, Xi'an Jiaotong University, Xi'an, 710049, China.

[4]Zernike Institute for Advanced Materials, University of Groningen, Nijenborgh 4, 9747 AG Groningen, the Netherlands

[5]Department of Physics & Materials Science, City University of Hong Kong, Kowloon, Hong Kong

[#]These authors contributed equally to this work.
*Corresponding authors.
E-mail: zhaoxinwu@mail.xjtu.edu.cn

**Abstract**

Utilization of triplet excitons, which generally emit poorly, is always fundamental to realize highly efficient organic light-emitting diodes (LEDs). While triplet harvest and energy transfer via electron exchange between triplet donor and acceptor are fully understood in doped organic phosphorescence and delayed fluorescence systems, the utilization and energy transfer of triplet excitons in quasi-two-dimensional (quasi-2D) perovskite are still ambiguous. Here, we use an orange-phosphorescence-emitting ultrathin organic layer to probe triplet behavior in the sky-blue-emitting quasi-2D perovskite. The delicate white LEDs architecture enables a carefully tailored Dexter-like energy-transfer mode that largely rescues the triplet excitons in quasi-2D perovskite. Our white organic-inorganic LEDs achieve maximum forward-viewing external quantum efficiency of 8.6% and luminance over 15000 cd m$^{-2}$, exhibiting a significant efficiency enhancement versus the corresponding sky-blue perovskite LED (4.6%). The efficient management of energy transfer between excitons in quasi-2D perovskite and Frenkel excitons in organic layer opens the door to fully utilizing excitons for white organic-inorganic LEDs.

**KEYWORDS**: Quasi-two-dimensional perovskite; White light-emitting diodes; Triplet excitons; High efficiency

# Introduction

Lead-halide perovskite materials, which have advantages such as high carrier mobility, high photoluminescence quantum efficiency, adjustable band gap, and low manufacturing cost,[1-3] show great potential in display and lighting. Taking advantage of 2D perovskite grains with higher exciton binding energy that can facilitate excitons formation and energy cascade to 3D grains for suppression of trap-assisted non-radiative recombination,[4-6] the EQEs of red and green quasi-two-dimensional (quasi-2D) perovskite LEDs[7, 8] have achieved over 20%, and blue quasi-2D perovskite LEDs[9-12] also break the EQE of 10% in the past a few years. Unlike traditional inorganic semiconductor material with high dielectric constant, the exciton binding energy in 2D and quasi-2D perovskites is relatively large (more than 100 meV) due to the dielectric and spatial confinement.[13] For example, the binding energy, splitting energy, and exchange energy of excitons in $(C_4H_9NH_3)_2PbBr_4$ have been determined to be 480, 70 and 31 meV resulting in a distinct singlet-triplet split (singlet state of 3.01 eV, and two triplet states of 2.982 eV and 2.988 eV respectively).[14, 15] In spite of these tremendous progress in quasi-2D perovskite LEDs, the utilization and energy transfer of triplets excitons in perovskite LEDs are seldom discussed.[16-18] However, triplet excitons are important as, in general, 75% of excitons formed during electronic excitation will be triplets according to the spin statistics.

Recently, triplet energy transfer (TET) across the lead halide perovskite nanocrystal/organic molecule interface with sufficient electronic coupling in hexane solution is reported,[19-21] whereas other studies on TET between perovskite and organic molecule in the films could hardly be found. The absence of investigation into the excitonic nature and excitons energy transfer also restricts the realization of efficient multi-color white perovskite LEDs involving energy transfer between excitons of different emitters like in white organic LEDs.[22-24] Although a white perovskite LED with maximum EQE of 6.5% is reported based on single-layer heterophase halide perovskites taking advantage of self-trapped emission,[25] recently, most of works focus on the combination of perovskite and other emitters with complementary colors, such as polymer, organic small molecule, and rare earth ions, by directly blending them[26-28] or adopting contacted dual emitter layer heterojunction[29, 30]. To date, however, these perovskite-based white LEDs have exhibited unsatisfactory EQE less than 2% obviously lower than the initial monochromatic perovskite counterparts without other emitters. In fact, directly blending or adopting contacted heterojunction would lead to a complex system, in which several effects are entangled with each other, such as energy transfer between perovskite and other emitters, defect introduction or passivation of perovskite by other emitters, and charge-carrier trapping in other emitters, as well as other non-radiative dissipation channels due to the dark triplet in introduced organic fluorescent emitters and concentration quenching. The mixture of effects severely impedes the understanding of mechanism in perovskite-based white LEDs, especially energy transfer associated with origin of perovskite excitons, and the further improvement of device efficiency.

Here, we report a concept of energy transfer management between quasi-2D perovskite and organic emitter. Through the adjustment of triplet energy level ($T_1$) and thickness of N-type spacer, Förster and Dexter-like energy transfer from the quasi-2D perovskite to the organic emitter can be controlled precisely. It is found that a Dexter-like triplet-triplet energy transfer can harvest the triplet excitons that tend to experience non-radiative transition and maintain the potential for unity excitons utilization efficiency. Efficient white LEDs with EQE= 8.6% is achieved, which is higher than the best EQEs of previous reports using all-perovskites or combined organic-perovskite emitting layer without any external out-coupling structure (Note S1, Table S1 and Figure S1, Supplemental Information).[25, 30]

## Results and discussion

The principle of excitons management in our white LEDs is illustrated in **Figure** 1. The inserted spacer plays a significant role in blocking holes and avoiding direct electron-hole recombination at ultrathin organic layer that leads to concentration quenching. The non-doped ultrathin (<1 nm) organic nanolayer, which can effectively reduce the cost and simplify the device structure as reported in the field of organic LEDs,[31-33] is proposed to harvest a part of energy from the perovskite layer. Excitons are formed in the perovskite layer, and the Förster-like energy transfer from perovskite to the ultrathin emitting layer can be regulated by altering the thickness of the inserted spacer, while a delicate Dexter transfer to channel the energy of triplet excitons in perovskite is expected by means of regulation of inserted spacer $T_1$ levels.

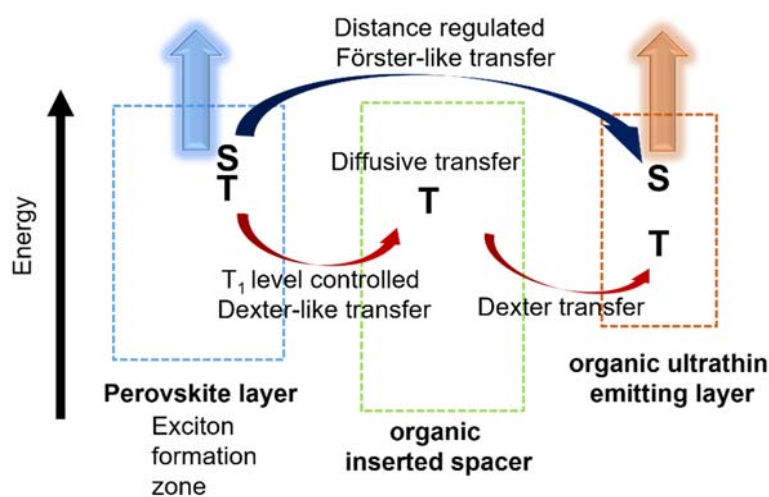

**Figure 1.** Proposed excitons management mechanisms in the white LEDs.

A mixture of 2D and three-dimensional (3D) perovskites is constructed with the aid of phenylethylammonium (PEA), and we selected the optimized quasi-2D $CsPbBr_{2.2}Cl_{0.8}$: PEABr (1:1 in precursor solution) perovskite film with photoluminescence quantum yield of 29% as smooth and pinhole-free blue emitting layer (Note S2 and Figure S2-7, Supplemental Information). The valance band of $CsPbBr_{2.2}Cl_{0.8}$: PEABr (1: 1) was measured by ultraviolet photoelectron spectroscopy (**Figure** 2d and Figure S8, Supplemental Information). Three commercial materials (1,3,5-triazine-2,4,6-triyl)tris(benzene-3,1-diyl))tris(diphenylphosphine oxide) (PO-T2T), 1, 3, 5-tris(Nphenylbenzimiazole-2-yl)benzene (TPBi), and bathophenanthroline (Bphen) as inserted spacer and electron transporting layer (ETL) with decreasing $T_1$ levels of 3.0 eV, 2.7 eV, and 2.5 eV were employed, respectively.[34-36] A commercial orange bis(4-phenylth-ieno[3,2-c]pyridine) (acetylacetonate)iridium(III) (PO-01) phosphorescent emitter with $T_1$ level of 2.2 eV, which can harvest both singlet and triplet excitons, is used to prepare ultrathin emitting nanolayer and examine triplet behavior in perovskite, whereas a conventional orange fluorescence dye 5,6,11,12-tetraphenylnaphthacene (rubrene) is utilized as comparative content.

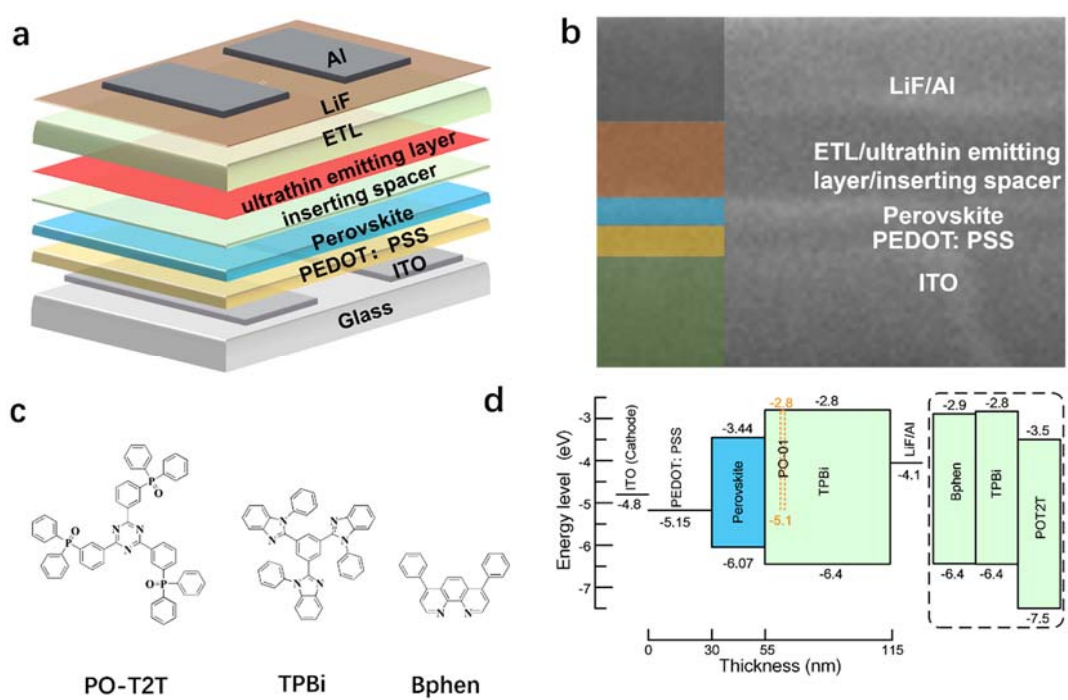

**Figure 2.** Schematic illustrations of the white LEDs device architecture: a) device structure of the proposed white LEDs. b) a cross-sectional SEM image of the white LED. c) chemical structure of the ETLs used in this paper with varying triplet energy levels. d) energy levels of the designed white LEDs.

Our designed white LEDs were fabricated with the configuration of indium tin oxide (ITO)/poly(3,4-ethylenedioxythiophene) polystyrene sulfonate (PEDOT: PSS) (~30 nm)/perovskite (~25 nm)/inserted spacer (x nm)/PO-01 (<1 nm)/ETL (60-x nm)/LiF (1 nm)/Al (100 nm) as sketched in Figure 2. First, we prepared a series of devices employing TPBi as inserted spacer and ETL (Figure S9 and S10, Supplemental Information). The diverse position and slight change in thickness of the ultrathin PO-01 layer have negligible impact on current density-voltage characteristics due to the ultrathin (<1 nm) thickness of PO-01 layer (Figure S9a and S10a, Supplemental Information). A maximum forward-viewing EQE of 8.6% and color rendering index (CRI) of 61 is achieved at a current density J= 3.3 mA cm$^{-2}$ using the optimized 7.5-nm-thick TPBi inserted spacer and 0.5-nm-thick PO-01 ultrathin layer, and the CRI could be enhanced to 80 with a maximum forward-viewing EQE of 7.2% and CIE coordinates of (0.40, 0.45) at a current density J= 1.7 mA cm$^{-2}$ via a device structure adopting double ultrathin organic layer (Note S3 and Figure S11, Supplemental Information). The maximum forward-viewing efficiency significantly drops to 2.5% in the device without TPBi inserted spacer due to concentration quenching in PO-01 ultrathin layer coming from direct formation of massive excitons at the organic PO-01 site. As shown in **Figure** 3b, when we change the inserted spacer and ETL to PO-T2T or Bphen with keeping the thickness parameter, the corresponding EQE is 5.6% or 0.9%.

To investigate the details of electroluminescence (EL) mechanism in the white LEDs, we should first determine the location of exciton recombination zone. We confirmed that a 7.5-nm-thick TPBi inserted spacer can completely block the holes (Note S4, Supplemental Information), thus we could draw a conclusion that the exciton recombination zone in the white LED employing a 7.5-nm-thick TPBi inserted spacer is located in the perovskite layer. Figure S14 shows the normalized EL spectra at different luminance of the white LED using a 7.5-nm-thick TPBi inserted spacer and an ultrathin PO-01 layer with thickness of 0.5 nm. Two obvious emission peaks at 484 nm and 564 nm are originated from perovskite and PO-01, respectively. As the driving voltage increases, the proportion of long wavelength orange-emitting component gradually reduces, indicating the gradual expansion of exciton recombination zone from perovskite/TPBi interface towards anode. The variation in EL spectra brings adjustable color temperature from a low color temperature of 2943 K, corresponding to Commission International de l'Eclairage (CIE) coordinates of (0.47, 0.48), to a high color temperature of 6013 K (CIE coordinates of (0.32, 0.39)) when varying the luminance from 100 to 10000 cd

m$^{-2}$ (Figure S14). This shows benefits toward physiologically-friendly solid-state lighting sources for more flexibility in diverse needs in different scenarios during the day or at night (Note S5, Supplemental Information).

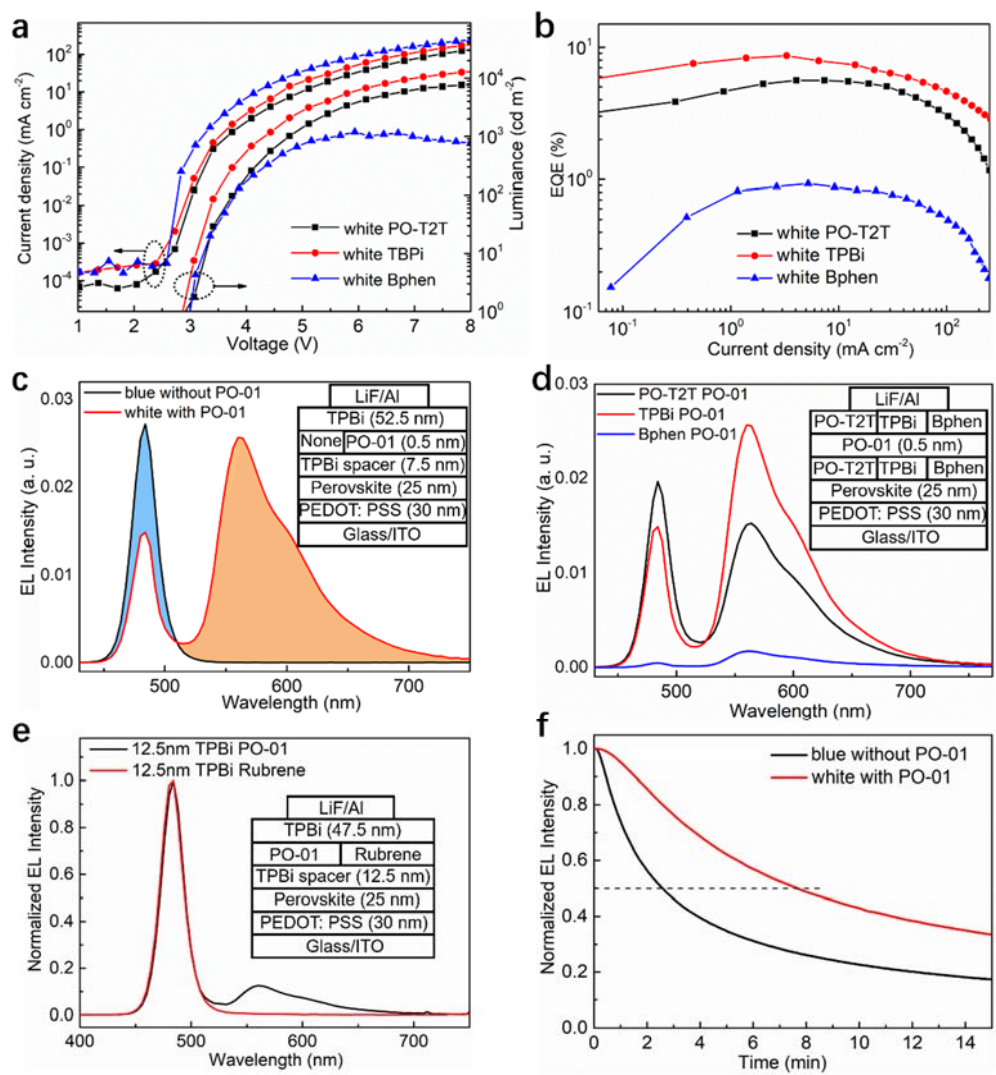

**Figure 3.** Device characteristics of the white LEDs: a) current density-luminance-voltage characteristics of the white LEDs with different ETLs. The device structure is ITO/PEDOT: PSS (~30 nm)/perovskite (~25 nm)/PO-T2T, TPBi, or Bphen (7.5 nm)/PO-01 (0.5 nm)/ PO-T2T, TPBi, or Bphen (52.5 nm)/LiF (1 nm)/Al (100 nm). b) EQE characteristics of the white LEDs with different ETLs. c) normalized EL spectra of the white device with a structure of ITO/PEDOT: PSS (~30 nm)/perovskite (~25 nm)/TPBi (7.5 nm)/PO-01 (0.5 nm)/ETL (52.5 nm)/LiF (1 nm)/Al (100 nm) at different luminance. c) EL spectra at a current density of 5 mA cm$^{-2}$ of the blue and white devices shown in the inset. The device structure is ITO/PEDOT: PSS (~30 nm)/perovskite (~25 nm)/TPBi (7.5 nm)/PO-01 (0.5 nm)/TPBi (52.5 nm)/LiF (1 nm)/Al (100 nm) or ITO/PEDOT: PSS (~30 nm)/perovskite (~25 nm)/TPBi (60 nm)/LiF (1 nm)/Al

(100 nm). d) EL spectra at a current density of 5 mA cm$^{-2}$ of three device structures shown in the inset. The device structure is The device structure is ITO/PEDOT: PSS (~30 nm)/perovskite (~25 nm)/PO-T2T or TPBi or Bphen (7.5 nm)/PO-01 (0.5 nm)/ PO-T2T or TPBi or Bphen (52.5 nm)/LiF (1 nm)/Al (100 nm). e) EL spectra at a current density of 5 mA cm$^{-2}$ of three device structures shown in the inset. The device structure is ITO/PEDOT: PSS (~30 nm)/perovskite (~25 nm)/TPBi (12.5 nm)/PO-01 (0.5 nm) or Rubrene (0.3 nm)/TPBi (47.5 nm)/LiF (1 nm)/Al (100 nm). f) The lifetime test of the blue and white devices at a current density of 10 mA cm$^{-2}$ in an N$_2$-filled glovebox. The device structure is ITO/PEDOT: PSS (~30 nm)/perovskite (~25 nm)/TPBi (7.5 nm)/PO-01 (0.5 nm)/TPBi (52.5 nm)/LiF (1 nm)/Al (100 nm) or ITO/PEDOT: PSS (~30 nm)/perovskite (~25 nm)/TPBi (60 nm)/LiF (1 nm)/Al (100 nm).

The corresponding sky-blue perovskite LEDs without ultrathin organic nanolayer were also fabricated to help understand our white LEDs. The forward-viewing peak EQEs of these blue perovskite LEDs employing TPBi, PO-T2T, and Bphen as ETL are 4.6%, 5.5%, and 0.1%, respectively (Note S6 and Figure S15, Supplemental Information). Remarkably, as depicted in Figure 3c, though EL intensity in blue region decreases, a drastic enhancement in orange region is observed in the white LEDs with TPBi spacer compared to blue LEDs using TPBi ETL at a same current density of 5 mA cm$^{-2}$. Analogous efficiency enhancement also appears in the Bphen-based white LEDs (Note S7, Supplemental Information), and the factor of out-coupling efficiency variation has been excluded for the efficiency enhancement (Note S8 and Figure S17, Supplemental Information). However, we have failed to see evident efficiency improvement in the white LEDs adopting PO-T2T spacer with a much high T$_1$ level. Figure 3d presents the EL spectra of three white LEDs at the same current density of 5 mA cm$^{-2}$. Weakest orange-emitting is observed in the white LEDs using Bphen spacer, and we can see a significant enhancement of orange emission derived from PO-01 in the device using TPBi spacer compared with the counterpart using PO-T2T, which performs better in blue LEDs. Furthermore, an inferior maximum forward-viewing EQE of 3.8% is observed in the optimized device (Figure S18 and S19, Supplemental Information) using a rubrene ultrathin layer and a 7.5-nm-thick-TPBi inserted spacer. These results suggest that energy transfer and utilization of triplet excitons might contribute to the superior performance in white LEDs with TPBi spacere except for Förster-like energy transfer between perovskite and PO-01. Figure 3e provides direct evidence for triplet excitons, where we observe an unambiguous orange emission peak in the device with phosphorescent PO-01 ultrathin layer in contrast to the rubrene-based device both using the 12.5-nm-thick-TPBi inserted spacer, thereby implying that triplet excitons do exist in the TPBi inserted spacer in view of the distinction for triplet harvest between phosphorescent PO-01 and fluorescence rubrene emitter. The device lifetime measurements of the white device using TPBi spacer and the corresponding blue device without ultrathin PO-01 organic nanolayer were operated under a constant current density of 10 mA cm$^{-2}$. As shown in Figure 3f, the white device achieves a half lifetime (50% loss of its initial EL intensity) of about 7.7 min, which is 3.1 times longer than that of the blue perovskite LEDs (around 2.5 min). The better device stability of the white LEDs is mainly attributed to the density reduction of perovskite excited states owing to energy transfer from perovskite to organic molecule.

To further understand the origin of triplet excitons in TPBi spacer, we prepared four film samples of $CsPbBr_{2.2}Cl_{0.8}$: PEABr (1:1) (perovskite), perovskite/PO-T2T, perovskite/TPBi, and perovskite/Bphen to investigate the exciton dynamics at the perovskite/organic interface. To enhance the contribution of interface excitons as similar in EL devices and still keep the smoothness of perovskite, the thickness of perovskite film is reduced to ~15 nm as much as possible. We carried out the transient PL and transient absorption (TA) measurements by using 400 nm excitation to ensure selective excitation of perovskite, in addition, Förster-like energy transfer from perovskite to organic layer is disallowed due to the lack of overlap between the absorption spectra of organic films and the PL spectra of perovskite (Figure S20 and S21, Supplemental Information). **Figure** 4a shows the PL decay of perovskite. The adding of TPBi layer has little influence on radiative recombination of perovskite, whereas the PL lifetime of perovskite sharply decreases by introducing Bphen layer, indicating obvious quenching of radiative electron-hole pairs. TA spectra of perovskite at different timescales in Figure 4b show two exciton bleach (XB) peaks at 454 nm and 479 nm, corresponding to n = 3 and n ≥ 4 phases in consistent with steady state absorption spectra in Figure S2 (Supplemental Information).[9] The TA kinetics in Figure 4c probed at 454 nm almost stay the same of all the films. Intriguingly, the TA kinetics probed at 479 nm show a gradually slower recovery as the $T_1$ level of organic layer decreases. Considering that the electron or hole transfer from perovskite to TPBi or Bphen molecule is energetically disallowed, the displayed unusual slower recovery is most consistent with a slow Dexter-like TET process from perovskite to TPBi or Bphen layer arising from small driving force as well as weak interfacial electronic coupling (Note S9 in Supplemental Information). Unfortunately, absorption features associated with the organic layer have not been observed in the TA difference spectra (Figure 4f and Figure S21, Supplemental Information).

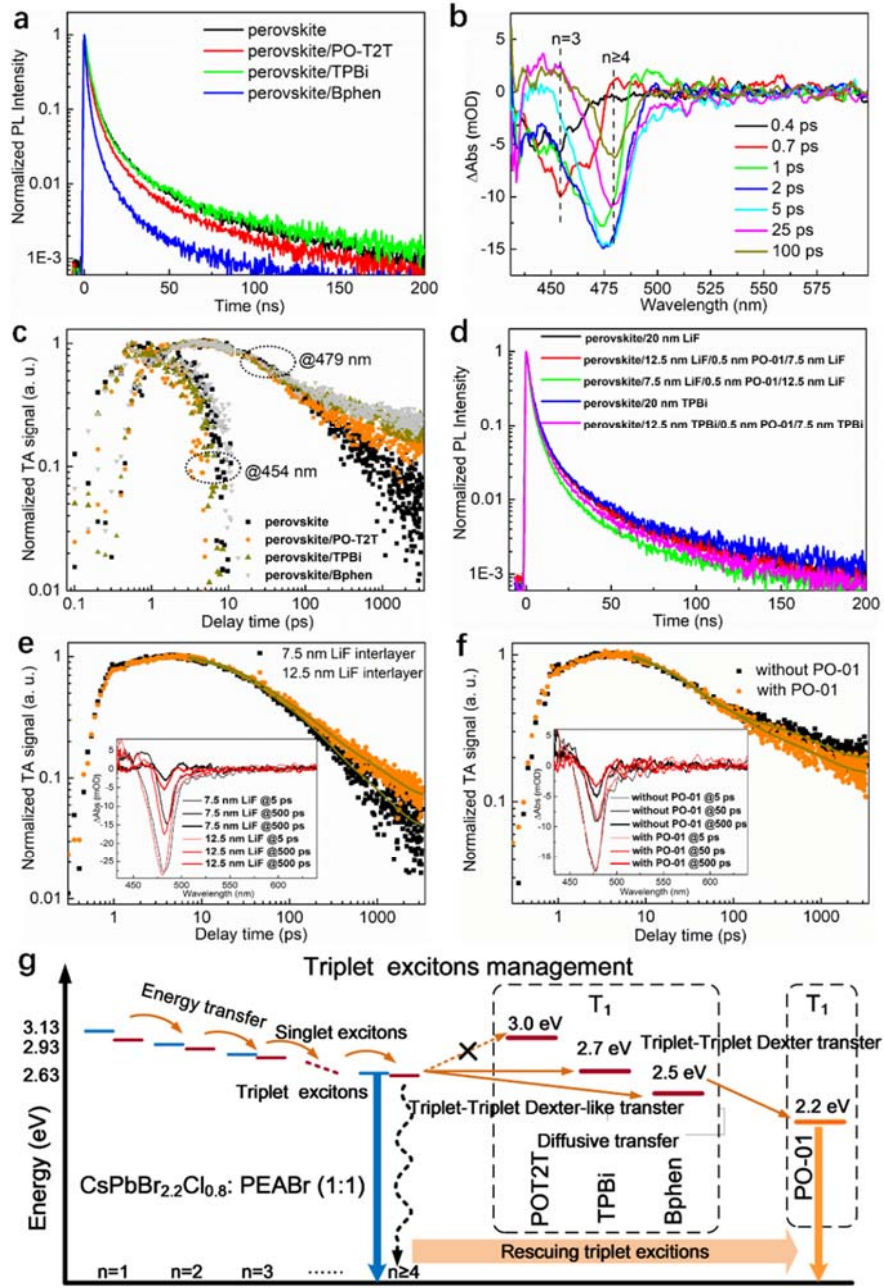

**Figure 4.** Proposed emission and energy transfer mechanism: a) PL decay measurements of CsPbBr$_{2.2}$Cl$_{0.8}$: PEABr (1:1) (perovskite ~15 nm), perovskite (~15 nm)/PO-T2T (20 nm), perovskite (~15 nm)/TPBi (20 nm), and perovskite (~15 nm)/Bphen (20 nm) films. b) transient absorption (TA) spectra of perovskite (~15 nm) film at selected time delays from 0.4 ps to 100 ps excited by a 400 nm laser pulse. c) TA kinetics probed at selected wavelengths (454 and 479 nm) for perovskite (~15 nm), perovskite (~15 nm)/PO-T2T (20 nm), perovskite (~15 nm)/TPBi (20 nm), and perovskite (~15 nm)/Bphen (20 nm) films. d) PL decay measurements of perovskite (~15 nm)/LiF (20 nm), perovskite (~15 nm)/LiF (12.5 nm)/PO-01 (0.5 nm)/LiF (7.5 nm), perovskite (~15 nm)/LiF (7.5 nm)/PO-01 (0.5 nm)/LiF (12.5 nm), perovskite (~15 nm)/TPBi (20 nm), and perovskite (~15 nm)/TPBi (12.5 nm)/PO-01 (0.5 nm)/TPBi (7.5 nm)
12

films. e) TA kinetics probed at 479 nm and spectra (inset) at different time delays following 400 nm excitation for perovskite (~15 nm)/LiF (7.5 nm)/PO-01 (0.5 nm)/LiF (12.5 nm) and perovskite (~15 nm)/LiF (12.5 nm)/PO-01 (0.5 nm)/LiF (7.5 nm) films. f) TA kinetics probed at 479 nm and spectra (inset) at different time delays following 400 nm excitation for perovskite (~15 nm)/TPBi (20 nm) and perovskite (~15 nm)/TPBi (12.5 nm)/PO-01 (0.5 nm)/TPBi (7.5 nm) films. g) Proposed management mechanism of triplet excitons.

We prepared another four film samples of perovskite/LiF (20 nm), perovskite/LiF (12.5 nm)/PO-01 (0.5 nm)/LiF (7.5 nm), perovskite/LiF (7.5 nm)/PO-01 (0.5 nm)/LiF (12.5 nm), and perovskite/TPBi (12.5 nm)/PO-01 (0.5 nm)/TPBi (7.5 nm) for PL delay and TA kinetics measurements in order to further understand the energy transfer process in white LEDs. As shown in Figure 4d, in the sample using a 7.5-nm-thick LiF insulating layer between perovskite and PO-01 to prevent TET, the perovskite PL lifetime reduction implies the occurrence of Förster-like energy transfer compared with the counterpart in the perovskite/LiF sample. The PL lifetime decrease disappears when the thickness of LiF layer is increased to 12.5 nm, indicating the absence of Förster-like energy transfer at this distance. Furthermore, we observed a faster recovery of perovskite XB probed at 479 nm in the sample using a 7.5-nm-thick LiF layer than the film using a 12.5-nm-thick LiF layer, and the kinetics discrepancy in TA spectra is visible after 50 ps following excitation (Figure 4e), thereby demonstrating a relatively fast Förster-like energy transfer process.

Interestingly, the perovskite/TPBi (12.5 nm)/PO-01 (0.5 nm)/TPBi (7.5 nm) film with an ultrathin PO-01 layer shows a slightly faster PL delay than the perovskite/TPBi (20 nm) sample as shown in Figure 4d. We could deduce that the PO-01 ultrathin layer, which can constantly absorb the triplet exciton energy from TPBi, may accelerate the triplet diffusive transfer in TPBi and facilitate the TET process at the perovskite/TPBi interface. A slight but clear acceleration of XB recovery probed at 479 nm is observed in the TA measurement with PO-01 ultrathin layer introduced at a distance of 12.5 nm from perovskite/TPBi interface after 500 ps following excitation, while the TA spectra almost stay the same after 5 and 50 ps following excitation, supporting the acceleration effect of the PO-01 ultrathin layer. These could be other evidence to verify that Dexter-like TET would occur in the perovskite/TPBi interface.

The proposed triplet excitons management mechanism is presented in Figure 4g. The well-designed Dexter-like TET channel instead of the Förster-like energy transfer between bright excitons would be really effective to extract the energy of triplet excitons in perovskite otherwise would tend to undergo non-radiative transition, leading to EQE enhancement in TPBi- and Bphen-based white LEDs compared to the corresponding blue devices. However, no EQE enhancement is observed in the device using PO-T2T spacer with high $T_1$ energy associated with blocking of Dexter-like TET process at the perovskite/PO-T2T interface. Triplet management is quite essential for achieving high performance devices employing perovskite and organic emitters.

## Conclusion

In summary, we report a new white LEDs architecture that utilizes organic phosphorescence and fluorescence emitters to probe the triplet behavior in quasi-2D perovskite. Triplet excitons in the sky-blue-emitting perovskite are largely rescued via a carefully tailored triplet-triplet energy transfer, and our white organic-inorganic LEDs achieve maximum forward-viewing external quantum efficiency as high as 8.6%. We demonstrate that triplet management is quite essential for achieving high performance devices employing perovskite and organic emitters. Our work thus inspires an idea of fully utilizing excitons for white organic-inorganic LEDs with the efficient management of energy transfer between excitons in quasi-2D perovskite and Frenkel excitons.

**Acknowledgements**

This work was financially supported by National Natural Science Foundation of China (Grant No. 62174128, 61805186, 12004298, 61705173, 61904145, 62027822, 11574248), the 111 Project (Grant No. B17035) and the China Postdoctoral Science Foundation (Grants No. 2018M633510).

**Author Contributions**

Yue Yu, Chenjing Zhao and Chenjing Zhao conceived and designed the experiments, including fabrication and analysis of the films and devices. Lihe Yan, Jinhai Si, Yuren Li, Yanmin Xu contributed to transient absorption measurement and analysis. Bo Jiao, Jingrui, and Jun Xi contributed to optical simulation. Yue Yu wrote the first version of the manuscript. Hua Dong, Jingfei Dai, and Fang Yuan contributed to the modification of manuscript. All authors discussed the results and contributed to the final version of the paper. Zhaoxin Wu supervised the project.

**Notes**: The authors declare no competing financial interest.

**Data and materials availability:** All data in the main text or the supplementary materials are available on request from the corresponding authors.